\documentclass{emulateapj}

\newcommand{\msun}{$M_\odot$}
\newcommand{\mbh}{$M_\mathrm{BH}$}
\newcommand{\mm}{$M_\mathrm{BH}$--$M_\mathrm{bulge}$}
\newcommand{\msig}{$M_\mathrm{BH}$--$\sigma_\mathrm{bulge}$}
\newcommand{\ml}{$M_\mathrm{BH}$--$L_\mathrm{bulge}$}
\newcommand{\mbulge}{$M_\mathrm{bulge}$}
\newcommand{\ms}{$M_\mathrm{*}$}
\newcommand{\mdm}{$M_\mathrm{DM}$}
\newcommand{\sbulge}{$\sigma_\mathrm{bulge}$}

\begin{document}
\title{The non-causal origin of the black hole--galaxy scaling relations}
\shorttitle{The non-causal origin of the black hole--galaxy scaling relations}

\author{
        Knud Jahnke\altaffilmark{1},
	Andrea V. Macci\`o\altaffilmark{1}
       }

\altaffiltext{1}{Max Planck Institute for Astronomy, K\"{o}nigstuhl 17, 69117
Heidelberg, Germany \email{jahnke@mpia.de, maccio@mpia.de}}

\begin{abstract}
  We show that the \mm\ scaling relations observed from the local to the
  high-$z$ Universe can be largely or even {\it entirely} explained by a
  non-causal origin, i.e.\ they do not imply the need for any physically
  coupled growth of black hole and bulge mass, for example through
  feedback by active galactic nuclei (AGN). { Provided some physics for
    the absolute normalisation,} the creation of the scaling relations can
  be fully explained by the hierarchical assembly of black hole and
  stellar mass through galaxy merging, from an initially uncorrelated
  distribution of BH and stellar masses in the early Universe. We show
  this with a suite of dark matter halo merger trees for which we make
  assumptions about (uncorrelated) black hole and stellar mass values at
  early cosmic times. We then follow the halos in the presence of global
  star formation and black hole accretion recipes that (i) work without
  any coupling of the two properties per individual galaxy and (ii)
  correctly reproduce the observed star formation and black hole accretion
  rate density in the Universe. With disk-to-bulge conversion in mergers
  included, our simulations even create the observed slope of $\sim$1.1
  for the \mm-relation at $z=0$.
  This also implies that AGN feedback is not a required (though still a
  possible) ingredient in galaxy evolution. In light of this, other
  mechanisms that can be invoked to truncate star formation in massive
  galaxies are equally justified.
\end{abstract}


\section{Introduction}\label{sec:intro}
About a decade ago tight correlations between galaxy properties and those
of central supermassive black holes (BHs) were empirically established: BH
masses scale with the luminosity, the mass and the velocity dispersion of
their host galaxies' bulges \citep{mago98, ferr00, gebh00, trem02, mclu02,
  marc03, haer04, guel09, jahn09b, merl10}. These correlations were
quickly interpreted to yield two important implications: (i) If these
correlations were to exist for a random set of galaxies, then every galaxy
should contain a supermassive BH \citep[e.g.][]{korm95} (ii) If every
galaxy contains a BH and given the observed scaling relations, the
evolution of galactic bulges and central BHs should be coupled by a
physical mechanism (the `co-evolution' picture).

The former statement appears to be true at least above some mass
threshold, and introduced a new ingredient, the central BH, in galaxy
evolution. The latter statement is based on the vast energy available from
accreting black holes, providing the easiest conceivable mechanism to
physically couple BH and bulge properties despite the difference in linear
scales. Coupling only a few percent of this energy in an ``AGN feedback''
\citep{silk98} to the gas within the galaxy would have vast implications
on the temperature and structure of the surrounding interstellar medium.
Ad hoc models \citep{gran04,dima05,crot06} were very successful in
generating feedback loops that involve the energy from AGNs for quenching
star formation (SF) and fueling of the AGN themselves. Different
incarnations of AGN feedback are in principle able to not only couple SF
and BH accretion, but to simultaneously fix a number of existing problems
in galaxy evolution, namely an overproduction of massive galaxies in
semi-analytic models as well as the inability to truncate SF fast enough
to reproduce the observed color--magnitude bimodality of galaxies
\citep{bald04}. This motivated the inclusion of AGN feedback in a
yet-to-be-determined form and physical description as the driving force
behind the BH--bulge scaling relations.

Although { the actual effectiveness and impact of} at least ``quasar
mode'' feedback models { is still unclear}, the interpretation of the
scaling relations as a physically coupled evolution is largely assumed to
be correct and continues to provide the basis for many studies. As an
example, the evolution of the scaling relations \citep{treu04, peng06a,
  peng06b, treu07, schr08, jahn09b, merl10, benn10, deca10b} is
investigated in order to constrain the physical drivers behind
co-evolution and the growth mechanisms of BHs.

\subsection{An alternative origin of the scaling relations}
\citet{peng07} demonstrated a thought experiment for the potential origin
of a \mm-relation without a physical coupling, but as the result of a
statistical convergence process. In short, he showed that in principle,
arbitrary distributions of \mbh/\ms-ratios in the early universe converge
towards a linear relation through the process of galaxy merging. In this
central-limit-theorem view, a large number of mergers will average out the
extreme values of \mbh/\ms\ towards the ensemble average. What was
deliberately left open in this experiment was whether there are enough
major galaxy mergers in the history of an actual galaxy in order to drive
this process far enough.\medskip

In this present study we pick up this thought by following realistic
ensembles of dark matter halos through cosmic time.  Our immediate aim is
to test whether the simple assembly of galaxies and their BHs according to
a $\Lambda$CDM merger tree is able to produce BH--galaxy scaling relations
from initial conditions at early times, where \mbh\ and \mbulge\ (or \ms)
were completely uncorrelated per individual galaxy. We circumvent the
inherent problems and degrees of freedom of a full semi-analytic model by
not trying to simultaneously solve the problem of SF truncation or correct
BH or stellar mass function, but restrict the question solely to the
genesis of the scaling relations. As an input for SF and BH accretion we
use observed relations, but prevent any recipes that im- or explicitly
couple BH and stellar mass growth {\em per individual galaxy}.


\section{Numerical simulations and merger trees}\label{sec:trees}
In this paper we use the Lagrangian code {\sc pinocchio} \citep{mona02} to
construct high resolution $\Lambda$CDM merger trees \citep[for a
  comparison between Nbody codes and {\sc pinocchio} see][]{li07}. The
simulation has a box size of 100 Mpc, and $1500^3$ particles, this ensures
a very high mass resolution, $m_p=1.01\times 10^7$ \msun.  The
construction of merger trees is straightforward with {\sc pinocchio}: the
code outputs a halo mass every time a merger occurs, i.e., when a halo
with more than 10 particles merges with another halo. From an initial
$6.5\times10^6$ halos, we receive a resulting sample of 10932 halos with
$M> 10^{11}$ \msun\ at $z=0$.

When two halos merge, the less massive one can either survive and continue
to orbit within the potential well of the larger halo until $z=0$, or
merge with the central object.  The averaging process described above will
only apply to this second category of halos (the ones that actually
merge), we then adopt the dynamical friction formula presented in
\citet{boyl08} to compute the fate of a halo.  { The orbital parameters
  of the halo are extracted from suitable distributions that reproduce the
  results of Nbody simulations as described in \citet{mona07}.}  If the
dynamical friction time is less that the Hubble time at that redshift, we
consider the { the halo to merger at a time $t=t_{dyn}$}, if it is
longer the satellite halo is removed from our catalog.  Each halo in our
sample at $z=0$ has formed by at least 200 mergers and the most massive
ones have had more than $5\times 10^4$ encounters.


\section{Creating scaling relations: averaging and mass function build up by
  hierarchical merging}
\label{sec:zero}
\begin{figure}
\includegraphics[width=\columnwidth]{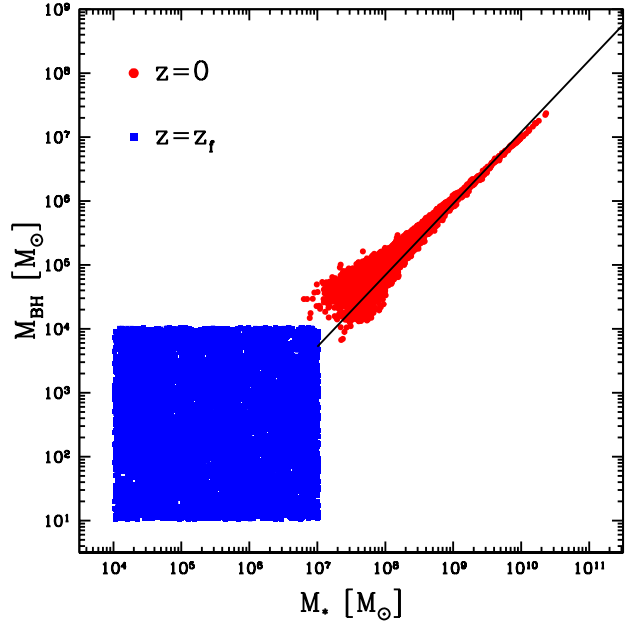}
\caption{ \label{fig:zero} 
Changes of \mbh\ vs.\ \ms\ from an initially uncorrelated (within 4~dex in
each parameter) distribution at high $z$ (blue points) to $z=0$ purely by
mass assembly along the merger trees, i.e.\ without SF, BH accretion and
disk-to-bulge mass conversion. A very tight correlation of slope 1.0 is
created by the merging, with smaller scatter for higher masses which
experienced more merger.
The black line is the observed local \mbh-\ms-relation from \citet{haer04}
with slope=1.12.
}
\end{figure}

\begin{figure*}
\begin{center}
\includegraphics[width=0.8\textwidth,clip]{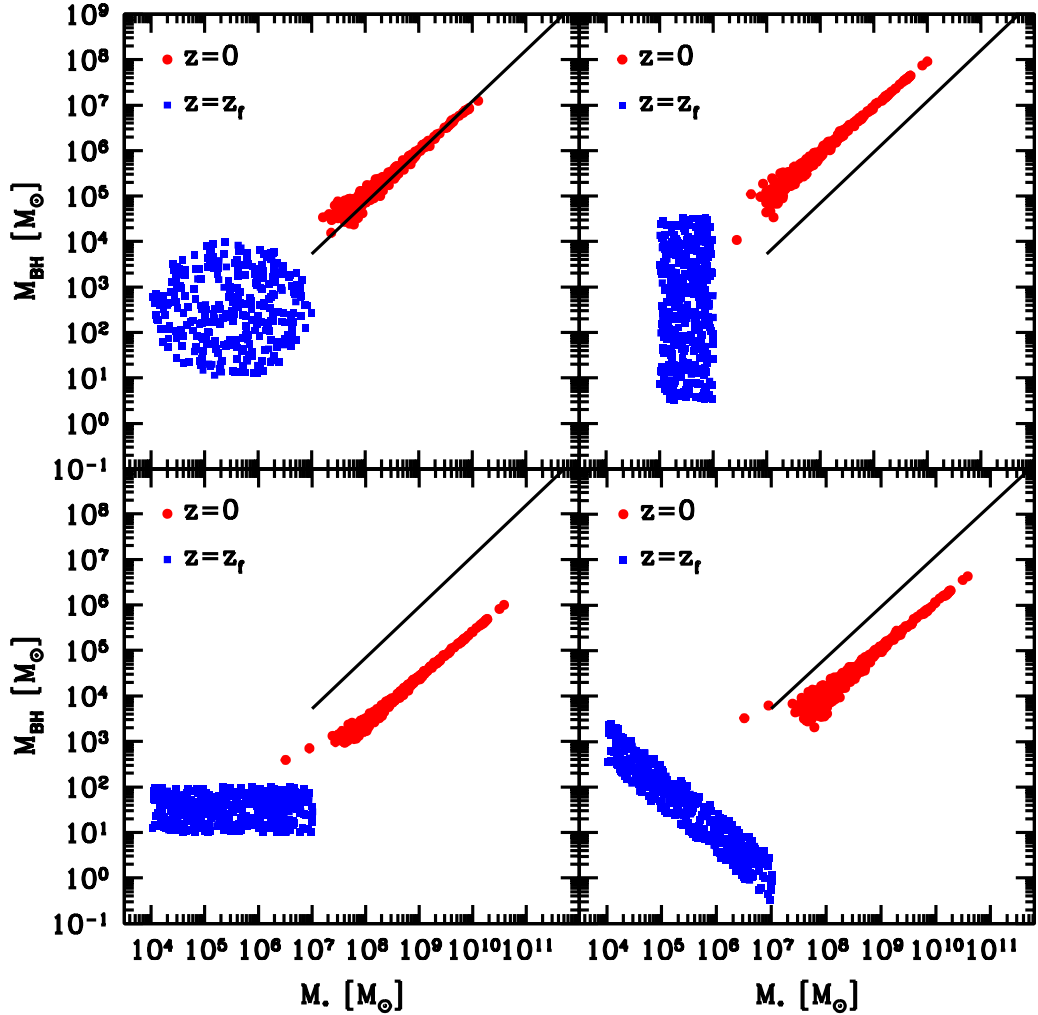}
\end{center}
\caption{ 
\label{fig:zero2} 
The scaling relations are produced independently of the initial conditions:
Shown are four vastly different initial conditions -- for a subset of our
halos for better visibility --, all leading to a slope=1 relation at $z=0$ with
similar scatter. The $z=0$ normalisation comes out differently since the
geometric mean of the initial
masses is different (flat distribution in logarith for both quantities).
Symbols and the line have the meaning as in Figure~\ref{fig:zero}.
}
\end{figure*}

{ The main message of this work is to demonstrate which effect merging over
  cosmic time has on an ensemble of halos with initially uncorrelated \mbh\
  and \ms\ values. These values change their distribution and converge
  towards a linear relation by $z=0$ -- in the absence of SF, BH accretion,
  disk-to-bulge conversion and hence any physical connection between the two
  masses.

{ For this task we follow dark matter halos through their assembly
  chain. We assign a stellar and a BH mass to each dark matter halo once
  its mass becomes larger than 10$^8$ \msun\footnote{We picked this mass
  since at lower masses halos likely did/do not form stars at all
  \citep{macc10}. The most massive progenitor of $z=0$ galaxies form
  according to this definition in the range $z=15-17$, while low mass
  satellites can form as late as $z\sim3$.}, the corresponding redshift in
the following is called $z_f$.}
}%
We set our initial guesses for
\ms\ and \mbh\ as a fixed fraction of the dark matter mass plus a (large)
random scatter.  We used $M_*/M_{\rm dm}=10^{-3}$ and $M_{\rm BH}/M_{\rm
  DM}=10^{-7}$ for the initial ratio; the scatter is taken from a
logarithmically flat distribution of { 3~dex for the two quantities} (blue
squares in Figs.~\ref{fig:zero} { and \ref{fig:zero2}}).

We have no knowledge of any realistic seed mass scatter, but take 4 orders
of magnitude variations in the \mbh/\ms-ratio as a proxy for
``uncorrelated''. { Empirical constraints on the possible parameter space
  for seed black hole mass do not seem to support seeds more massive than
  $\sim$$10^5$ \msun \citep{volo09}, towards lower masses a few solar mass
  BHs are clearly being produced by stars. Whether this matches the true
  distribution of seed masses is not important, but by simply taking a
  large range that is currently not ruled out represents a rather
  conservative starting point for our demonstration.}  Halos are then
propagated along the merger tree to $z=0$ (the red points in
Figs. \ref{fig:zero} { and \ref{fig:zero2}}). When two halos merge
according to our dynamical friction formula, we set the resulting stellar
and BH masses equal to the sum of the individual masses before the merger
\citep{volo03}. { The final mass in \mbh\ and \ms\ as well as the
  corresponding normalization is determined simply by the sum of the
  individual halos contributing to a final halo.}

Fig.~\ref{fig:zero} shows that the hierarchical formation of galaxies provides
a strong inherent driver from the uncorrelated initial distribution to a
linear relation\footnote{The convergence is in fact too strong (see next
  sections), as the scaling relation it produces by $z=0$ is much tighter than
  the observed 0.3~dex scatter. In principle the scatter has a $\sqrt{N}$
  dependency on the number $N$ of merger generations, but the relation gets
  complicated by the different masses of the merging components and different
  merging times across the tree.}. { This effect is independent of the
  chosen initial conditions as Figure~\ref{fig:zero2} demonstrates, where
  completely different initial conditions result in a relation with the same
  slope and very similar scatter.}

{ This experiment shows that the dominating structural parts of the
  observed \mbh-\ms-scaling relation -- i.e.\ (1) the existence of such a
  correlation, (2) that it extends over several orders of magnitude in mass,
  (3) the fact that the slope is near unity, and (4) an increasing scatter to
  lower masses -- can be explained by this physics-, feedback- and
  coupling-free process. A slope$\sim$1 scaling relation does not need any
  physical interaction of galaxy and black hole. In the next sections we will
  show that this holds also when adding ``2nd order'' effects like actual star
  formation and black hole accretion, as well as disk--bulge conversion.}


\section{Adding star formation, black hole accretion, 
and disk-to-bulge conversion}\label{sec:one}
\begin{figure}
\includegraphics[width=\columnwidth]{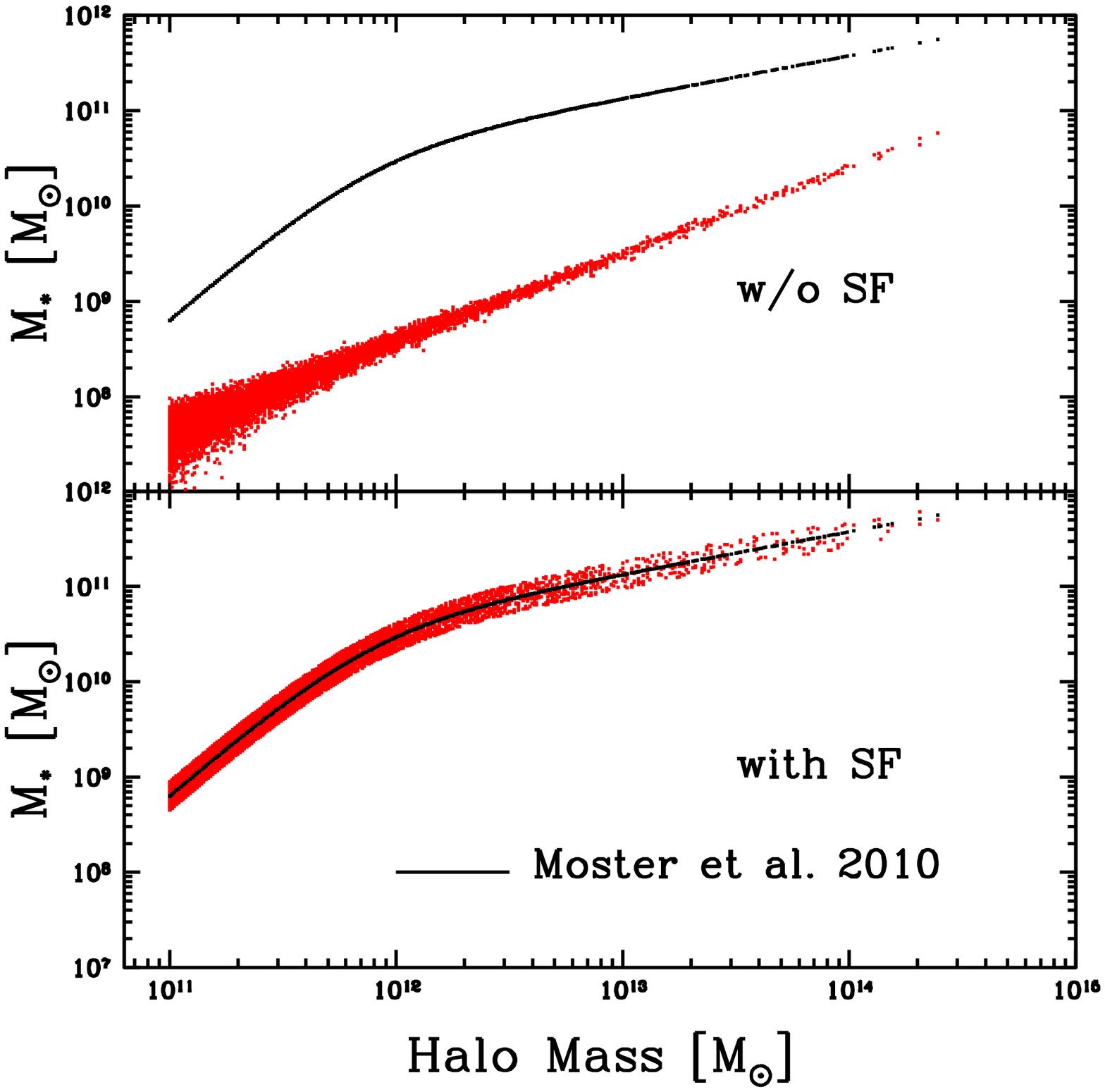}
\caption{
The relation between \ms\ and \mdm\ at $z=0$. The black line shows the
predictions from \citet{most10}, the red point the results from our
simulations. Upper: Only merging but no SF taken into account. Lower:
With our SF recipe implemented.
}
\label{fig:ben} 
\end{figure}

{ We so far demonstrated that merging alone is the basic mechanism to
  create a \mbh--\ms\ scaling relation. However we need to add a number of
  ingredients to our model: Placing all mass already at high redshift is not
  conservative, since all of stellar and BH mass is subject to the full merger
  averaging process. In the actual universe we know that the majority of BH
  and stellar mass in the universe was created after $z\sim 6$ \citep{solt82,
    hopk06}, and hence experiences less merger generations. Moreover pure
  merging produces a monotonic relation between \mdm\ and \ms\ as shown in the
  upper panel of Figure \ref{fig:ben}, which is at odds from empirical
  results. Since SF and BH accretion density depend on redshift, we will add
  the approximate right amount of SF and BH growth at the right cosmic times,
  and thus at the right place in time with respect to the merger cascade. The
  goal of this excercise is not to create a full semi-analytic model of galaxy
  formation, but to test which effect realistic assumptions about mass growth
  have on the resulting \mbh--\ms\ scaling relation.}

To construct SF and BH accretion recipes we will use three observed relations
as input: (i) The halo occupation distribution, i.e.\ the relation between
dark matter halo mass \mdm\ and \ms\ \citep{most10}, (ii) the Lilly-Madau
relation for the evolution of SF rate \citep{hopk06}, and (iii) the evolution
of bolometric luminosity of AGN \citep{hopk07}.

We want to explicitly note that by using these relations, we do not force
any coupling of \mbh\ and \ms\ or \mbulge\ {\em per individual galaxy},
but all recipes only relate to ensemble averages. Any potential implicit
couplings only act on the ensemble and not on an individual galaxy, hence
they can not induce a correlation in the (originally uncorrelated) data
points.
 
\subsection{Star Formation}\label{sec:sf}
{ Up to now, at redshift $z=0$ the stellar mass of our galaxies is simply
  given by the sum of all stellar masses of its $j$-progenitors $M^0_*(i)=\sum
  M_*(z_{f})(j)$, masses that as explained were originally drawn from a random
  distribution. As shown in the upper panel of Fig.~\ref{fig:ben}, for a given
  halo mass the stellar mass obtained in this way is too low when compared to
  the empirical expectations from Halo Occupation Distribution (HOD) models
  \citep[e.g.][]{most10}. This is because the total galaxy is more than the
  sum of its seeds and we have neglegted star formation so far. On the other
  hand, Fig.~\ref{fig:ben} tells us exactly how much stellar mass each halo is
  missing. Hence we take this constraint, the HOD results, to fix the stellar
  mass produced throught star formation:
\begin{equation}
M^*_{SF}(i)=M^*(M_{dm}(i))-M^*_0(i)
\label{eq:sfr1}
\end{equation}
where $M^*(M_{dm}(i))$ is the expected stellar mass for the $i$-th halo with
dark matter mass $M_{dm}(i)$ as predicted by the HOD model presented in
\citet{most10}.

Now we need to distribute this stellar mass from SF along time, i.e.\ among
all progenitors of galaxy $i$ along the merger tree. We do that according to
the following formula that gives the stellar mass produced through SF for the
$j$-th halo in the merger tree of final $i$ halo:
\begin{equation}
M^{SF}_j = A \times M_{*}^q(j) \times LT(j) \times f(z_f(j),z_m(j)).
\label{eq:sfr2}
\end{equation}
The constant $A$ is fixed by the requirement that $M^{SF}(i)=\sum_j M^{SF}_j$
and is the same for all progenitors.  The time $LT$ is the lifetime of a halo,
defined as time between the formation redshift ($z_f$: when
$M_\mathrm{DM}>10^8$ \msun) and the moment $z_m$, when it merges with a more
massive halo, which is not necessarily the main branch of the merger
tree. With this definition we assume that galaxies are able to actively form
stars only when they are the central object within their host halo.

The function $f$ is used to give different weights to the life time $LT$
at high ad low redshift, in this way, for a given life time, a galaxy will
produce more (less) stars at high (low) redshift, according to the
Lilly-Madau plot\footnote{ We note that we do not include a different
  shape of this star formation history as a function of mass. As the
  Appendix shows, this will not have any effect on the results}.  We
define $f(z_1,z_2)$ as the integral between $z_1$ and $z_2$ of the assumed
star formation rate (SFR):
\begin{equation}
f(z_1,z_2) = \int_{z_1}^{z_2} \rm SFR(z) \rm d z.
\end{equation}
In our reference model we assume for the redshift evolution of the SFR the
functional form suggested by \cite{hopk06}, namely results listed in
Table~\ref{tab:mods} for a modified Salpeter IMF \citep[see][for more
details]{hopk06}.  Finally the factor $ M_{*}^q(i)$ takes care of the observed
mass-dependence of specific star formation rates and we fixed the exponent
$q=0.8$ \citep[e.g.][]{dadd07,bouc09}. In the Appendix~\ref{sec:test} we will
present results for different choices for $q$ and $SFR(z)$.

Let us summerize one more time our parametrization for star formation: When
the $j$ halo appears at $z_f(j)$ it gets an initial stellar mass
($M_*(z_{f})(j)$) from a random distribution as described in
Section~\ref{sec:zero}. Then it will ``produce'' its own stars ($M^{SF}_j$)
until it will be accreted onto, and become part of, a more massive
halo. During its lifetime it will also accrete stellar mass from merging
(lower) mass haloes.  If halo $j$ will merge with the central galaxy it will
add a fraction of its stellar mass to the bulge of the central galaxy as
described below in Section \ref{sec:bd}. If halo $j$ will merge with a more
massive halo ($k$) before merging with the central halo it will give all its
stellar mass to halo $k$ and cease to exist as a halo of its own.
}

\subsection{Disk to bulge conversion}
\label{sec:bd}
We assume that all stellar mass produced through star formation will occur in
the disk component of each halo and then apply a recipe to convert part of
this disk mass to bulge mass as a consequence of mergers.

The amount of disk-to-bulge mass conversion depends on a multitude of
parameters as mass ratio, gas fraction, and orbital parameters of the merger,
which are impossible to implement in our context. Instead we follow a simpler
recipe inspired by the numerical results of \citet{hopk09}, solely depending
on the stellar mass ratio of the two merging partners: (i) Bulge mass of main
halo and satellite will just be co-added, (ii) the disk mass of the satellite
fully goes into the resulting bulge, and (iii) a fraction of the main halo
disk, { directly} proportional to the mass ratio, also gets converted into
bulge mass.  With this recipe we are able to approximately reproduce the ratio
of bulge to total mass observed in the local universe.

\subsection{Black hole accretion}
\label{sec:bha}
Since the mechanisms of BH accretion continue to be unclear, we assume a
simple recipe; the BH will double its mass in a stochastic way on a
characteristic time scale $\tau$.  { This will happen for all $j$
  progenitors of halo $i$. Similar to what we have done for star formation, we
  link the number of doublings of a given halo to its life time and we weight
  this time with a function $g$ similar to the function $f$ in
  Section~\ref{sec:sf}.

  In practice the number of mass doublings of the $j$-th black hole in the $i$
  merger tree is given by the expression:
\begin{equation}
N_{\mathrm{doub,}j} = LT(j) \times g(z_f(j),z_m(j)) /\tau,
\label{eq:bha}
\end{equation}
Where $LT$ has the same meaning as in Eq.~\ref{eq:sfr2}. For black hole
accretion we choose as the weighting function $g(z_1:z_2)$ the integral
between $z_1$ and $z_2$ of the bolometric luminosity of AGN ($AGN_L(z)$):
\begin{equation}
g(z_1,z_2) = \int_{z_1}^{z_2} \rm AGN_L(z) {\rm d} z.
\end{equation}

In our reference model the redshift evolution of the AGN bolometric luminosity
is modeled with a double power law aimed to reproduce the results of
\citet[Figure 8]{hopk07}:
\begin{equation}
\log\left(\frac{AGN_L(z)}{L_\odot \mathrm{Mpc}^{-3}}\right) = \left\{
\begin{array}{ll}
 2.02 \cdot \log(z) + 7.83 & \textrm{for $z<1.7$} \\
 -2.09 \cdot \log(z) + 8.78 & \textrm{for $z\ge 1.7$}
\end{array} \right. 
\end{equation}
Similarly to the $f$ function, $g$ can be used to allow for higher accretion
rates at high redshift compared to low redshift for a fixed halo life time.

The characteristic time $\tau$ is chosen in order to
match the normalization of the observed \mbh--\mbulge-scaling relation at
$z=0$ for $\log(M_\mathrm{BH})=7$ and we obtained $\tau=1.9 \times 10^9$ Gyrs.
As a stochastic element to BH growth we cast for each of the
$N_{\mathrm{doub,}j}$ events a random number from a flat distributing in the
range [0:1] and effectively double the BH mass only if this random number is
$> 0.5$. For the main branch the number of doublings is of the order of 6--8,
while it is in the range 0--3 in the other branches of the tree. In the
Appendix~\ref{sec:test} we will present results for different choices of the
parameters involved in equation~\ref{eq:bha}.  }

\begin{figure*}
\includegraphics[width=0.48\textwidth]{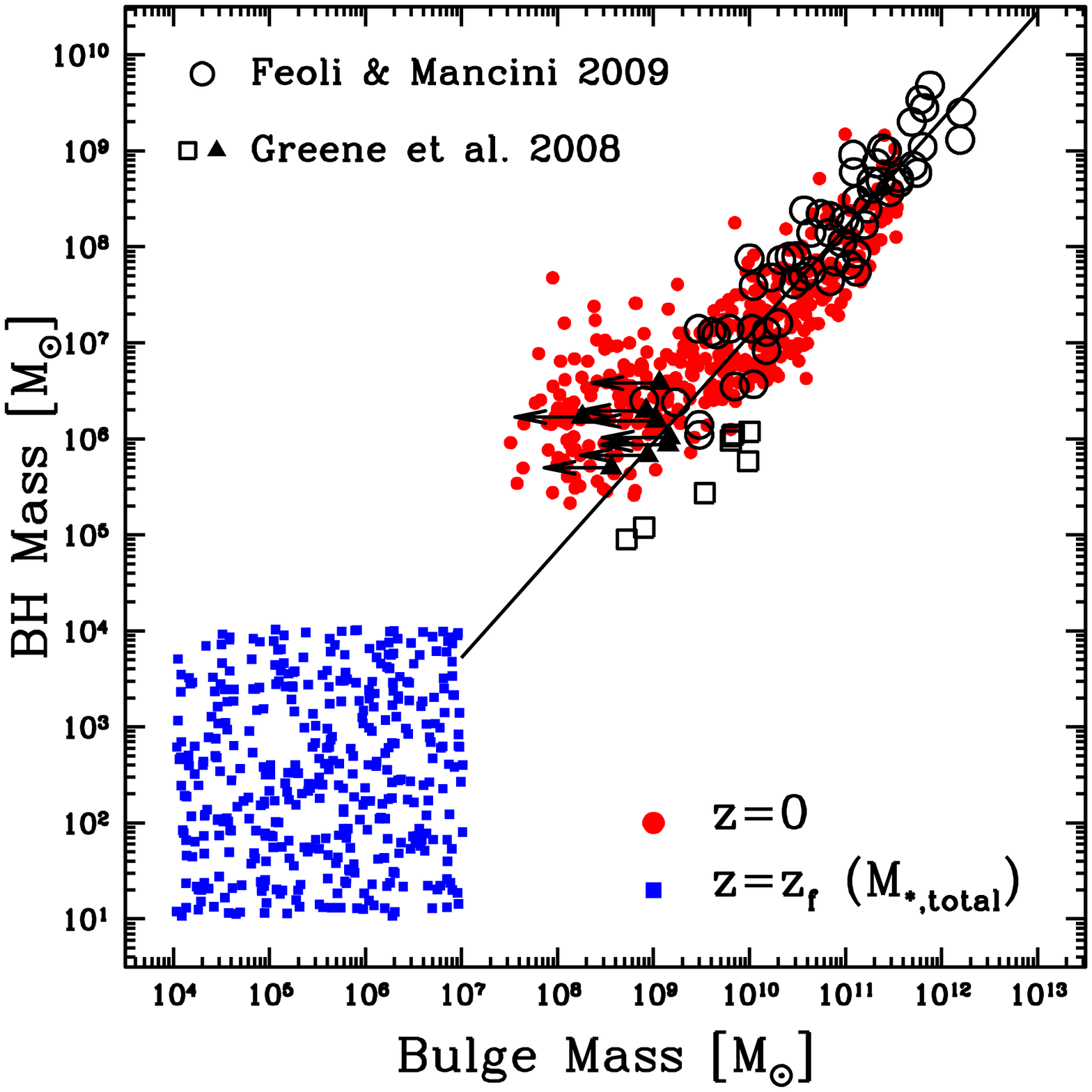}
\hfill\includegraphics[width=0.48\textwidth]{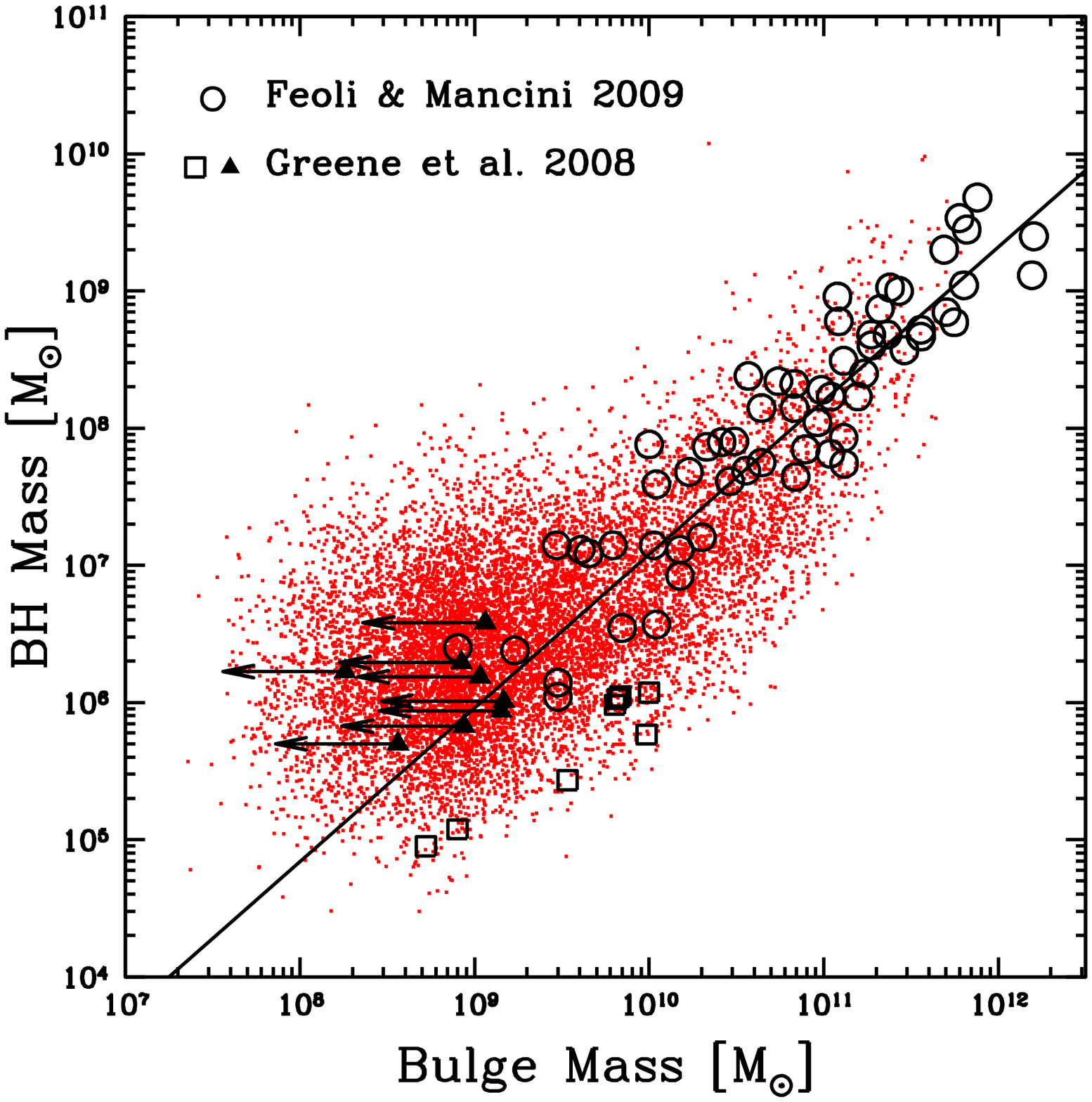}
\caption{ 
Left: Initial uncorrelated high redshift seeds for \ms\ and \mbh\ (blue
filled squares) and resulting $z=0$ \mm\ scaling relation for a subset of
400 randomly selected merger trees (red points), compared to the observed
local relation in black, including the compilation from
\citep[circles]{feol09}, low-mass spheroids (open squares) and upper
limits for spiral bulges (triangles) from \citet{gree08}. The solid line
is the linear fit by \citet{haer04} with a slope of 1.12.
Right: The full set of resulting 10932 galaxies at $z=0$ with the low--z
data overplotted.
}
\label{fig:MbMbh1} 
\end{figure*}

\subsection{Resulting scaling relations}
\label{sec:results}
The \mbulge--\mbh-distribution at $z=0$ with the above receipes added is
shown in Figure~\ref{fig:MbMbh1}, with the observed values
overplotted. Compared to the pure merger assembly in
Figures~\ref{fig:zero} and \ref{fig:zero2} we note a substantially
increased scatter -- closer to the observed -- and a steeper than linear
relation. Still, despite both the shift of SF and BH growth to later times
as well as the random parts of SF and BH accretion, a clear correlation is
produced. The effects of SF, BH accretion and disk-to-bulge conversion do
not destroy the correlations but only induce a ``2nd order'' modification
of them. Over time a mass function is built up and the scatter decreases
substantially from the initial 4~dex, particularly for the
$\log(M_\mathrm{bulge})>10$ regime.

The simulated relation is very similar to the observed points, it
reproduces the slope $>$1 almost perfectly, even the fact that at the high
mass end the observed points lie above the mean slope -- and this without
adjustable parameters beyond normalization. { Also, the higher scatter
  at low masses can be seen in both simulations and observations, the
  ``cloud'' above the $z=0$ relation in our model data represents
  disk galaxies with small bulge masses as observed by \citet{gree08}. As
  can be seen in Figure~\ref{fig:mstellar} in the Appendix, these objects
  would still lie on the local mass scaling relation, but with their
  total and not their bulge mass.}

We want to stress that the results in Figure~\ref{fig:MbMbh1} do not
depend on our parametrization of SF rate or BH accretion. For example
changing the functional form of $f$ or $g$ in Eqs.~\ref{eq:sfr2} and
\ref{eq:bha} only marginally affects the scatter of the simulated
\mm--relation and leaves the slope unchanged. The same is true for the
other parameters described in the previous sections, see the Appendix for
details. There is a sole exception to this, the assumed initial seeding
masses: The larger the initial \mbh\ and \ms, the less mass has to be
created by SF and BH accretion. { In this way less mass is
  entering the halos at later times and more mass is subject to the full
  cascade of mergers, which will lead} to a smaller scatter in the scaling
relations at $z=0$.


\section{Discussion}\label{sec:discussion}
{ We showed above that all basic properties of the BH--bulge mass scaling
  relation in the local Universe -- a relation between properties of
  individual galaxies -- are produced naturally by the merger-driven assembly
  of bulge and BH mass, and without any coupling of SF and BH mass growth per
  individual galaxy.

  The convergence power of galaxy merging is very strong for a realistic halo
  merger history, even with a correct placement of SF and BH accretion along
  cosmic time. This means that the mechanism \citet{peng07} sketched in his
  thought experiment works also in a realistic Universe -- there is enough
  merging occurring in the universe and hence (most of) the scaling relations
  can be entirely explained without any physical mechanisms that directly
  couples \mbulge\ and \mbh\ growth for a given object.
}

\subsection{Implicit coupling of black hole and galaxy?}
{ The scaling relations are produced naturally in our toy model -- but does
  this preclude any implicit coupling of \mbh\ and \ms\ by design? The most
  obvious features to be discussed in this respect are (a) the shape of the
  halo occupation distribution and (b) the question, which mechanism
  determines the relative value of BH accretion to SF, hence the absolute
  normalisation of the \mm-relation.

  (a) HOD shape: The HOD was empirically inferred and shows that the ratio of
  stellar to dark matter mass is not constant but is a function of mass
  itself. Towards the massive end stellar mass does not increase in parallel
  to DM, star formation appears suppressed. Its impact on the \mm-relation is
  the slight curvature in Figure~\ref{fig:MbMbh1} with the noticable upturn at
  the massive end -- consistent with the observations. The HOD shape
  represents a long known feature of galaxy formation and was ``fixed'' in
  models by introduction of a quenching mechanism, suppressing SF above some
  mass, often by inclusion of AGN feedback receipes.

  In our toy model however, we do not make specific assumptions about which
  mechanism produces the HOD we use as a constraint. We do not tie BH
  accretion or a merger event to the suppression of SF -- it can be suppressed
  by any mechanism, e.g.\ a modified SN feedback receipe or gravitational
  heating by infalling clumps of matter. The latter mechanisms are completely
  independent of BH accretion and, while modifying the HOD, by nature can not
  have an impact on the \mm-relation. Even if AGN feedback was the source for
  shaping the HOD, this would only be the cause of the second order shape
  deviation from a linear slope, not the existence of the \mm-relation itself.

  (b) Absolute normalisation: Our initial model just propagates the (rather ad
  hoc) seed masses in stars and BH to $z=0$, i.e.\ the normalisation of the
  \mm-relation at $z=0$ is directly set by the ensemble mean ratio of seed
  masses. Since most of stellar and BH mass in reality is produced by SF and
  BH accretion later on, the high-$z$ ratio is in fact unconnected to the
  $z=0$ normalisation.

  In our simulation we set the normalisation by requesting a match of our
  simulation results with the empirical \mm-relation at
  \ms=$10^{10}$\msun. Arguments have been brought forward that the actual
  normalisation must be the result of a regulatory feeback loop involving
  BH accretion and SF. This is an attractive scenario, as it would explain
  both the creation of the scaling relations as well as their
  normalisation. Models of this kind have been implemented in several
  semianalytic models of galaxy formation, in all cases with free
  parameters that actually control the absolute normalisation of the
  resulting scaling relations, set to ad hoc values to again match this
  and other observations. Since we show in this paper that (most aspects
  of) the \mm-scaling relations are created automatically by hierarchical
  assembly, these feedback models actually { seem to}  achieve too much --
  the creation of a certain \mbh/\ms\ ratio for {\em each galaxy} and at
  all times, which as a conspiracy would come on top of the formation path
  of the scaling relations demonstrated here.

  This said, we want to sketch the outline of an alternative scenario, that
  could well be responsible for the absolute normalisation but has not yet
  been explored: The main ingredients for both SF and BH accretion are (i)
  gas, and (ii) a trigger to form stars or to bring down gas to the BH. For
  both stars and BHs the amount of growth is basically a product of the two
  ingredients. At early times gas was ample and the number of both galaxy
  mergers and gas disk instabilities was high until the peak of activity
  around $z=2$. The triggering mechanisms subsequently decreased with the
  decreasing number of mergers and reduced gas reservoirs, either in number or
  duration or both. If both SF and BH accretion were to be ruled by a set of
  random triggering mechanisms and the specific gas fraction in a galaxy, then
  very high-$z$ galaxies might exhibit a strong variance in their
  \mbh/\ms-distribution as well as in their (instantaneous and also time
  averaged) BH accretion over SF ratio, with a dependency on the actual total
  mass, morphology, gas mass, trigger type, environment, etc. However, as a
  cosmic mean there will be a {\em global} \mbh/\ms\ value, just as an average
  over the efficiency of the spectrum of random triggering mechanisms and
  realised conditions to produce new stars or BH mass -- a number which could
  also change with time.

  The hierarchical assembly of galaxies and its averaging mechanism now
  relieves us from having to search for regulatory mechanism {\em per galaxy},
  since with each galaxy merger the spectrum of \mbh/\ms\ values will
  increasingly average out. The fact that the slowly starting depletion of gas
  reservoirs in galaxies at $z\sim2$ is accompanied with, and not preceded by,
  a slowly decreasing merger rate has the consequence that at higher redshifts
  there were enough mergers to average out the extreme \mbh/\ms\ systems -- at
  lower $z$, with a decaying gas reservoir and merger rate and the transition
  to a ``secular universe'' \citep[see e.g.][]{cist11}, the prerequisites for
  producing new extreme values become less and less frequently fulfilled. The
  ensemble converges towards the observed \mm-relation at $z=0$.

  Recently, first nested multi-scale simulations of gas inflow into the very
  centers of galaxies have been successful \citep{hopk10} and give a first
  impression of how in principle random instabilities, strongly depending on
  the actual conditions in the galaxy, can create gas inflow into the very
  center and the fuelling of either BH, SF or both. It is mechanisms of this
  kind that will create a certain mean value of \mbh/\ms\ averaged over all
  galaxies, which can be vastly different for an individual galaxy. How
  different is so far unclear, and whether this mechanism for BH fuelling
  precludes ``runaway'' growth of BHs to 100- or 1000-fold in a single
  instance, though unlikely, needs to be seen. All of this can in principle be
  realized without any AGN feedback -- though it does not rule it out --, and
  has the freedom to have a mass- or environmental-dependent efficiency
  component, as we also observe a non-unity slope of the $z=0$ scaling
  relation.

  Coming back to the initial question, we conclude that there is no necessity
  that our toy model makes any implicit assumptions about a physical
  \mbh--\ms-coupling.
}

\subsection{Further consequences of this mechanism}
{ 

  Our mechanism is also able to explain other observational results that are
  often used as evidence to support the picture of agn feedback.

  (a) One of them is the potentially different \mm-relations in galaxies with
  classical and pseudo-bulges \citep{gree08,gado09}. Pseudo-bulged are thought
  to be formed through secular processes, rather than major merging
  \citep{korm04}. This has the immediate implication that the bulge has taken
  a different long- or  mid-term assembly route compared to the BH,
  hence it is actually not expected that galaxies with pseudo-bulges obey the
  same \mm-relation than those with classical ones.

  (b) \citet{hopk09} found that the stellar mass inside a very small radius
  near the black hole \ms($<R$), comparable to the BH's sphere of influence,
  shows a much larger scatter than the scatter in \mbh/\ms. They argue that
  this is an indication that gas was transported to near the BH, to form said
  stars, but that this had apparently no impact on the scatter in \mbh, hence
  a self regulating mechanism should have been at work.

  With our model also this can be simply explained: The BH and bulge of a
  galaxy take part in the same long-term assembly and averaging cascade, hence
  their values correlate and scatter in their ratio is small. The central
  density of stars on the other hand does not. It is governed by more short
  term gas inflow, star formation and redistribution mechanisms, hence it is
  expected that it does not correlate well with the overall mass of the bulge
  or the BH. 

  (c) Following this line of argument, we in princple expect a correlation
  with \mbh\ for any (mass) parameter that is subject to the same
  $\Lambda$CDM assembly chain. This includes the halo mass, to some extent
  the total mass of the galaxy { (for both see Appendix~\ref{sec:stellardm}
  and Figure~\ref{fig:mstellar})}, but also e.g.\ the total mass of globular
  clusters in a galaxy, which recently has been found empirically by
  \citet{burk10}.

  (d) On the other hand, our model is too basic to be able to reproduce higher
  order effects. A number of studies \citep{alle07, hopk07, feol09, feol10}
  have suggested that the most fundamental relation with \mbh\ is neither
  \mbulge\ nor \sbulge\ but that rather galaxy binding energy or potential
  well depth. 

  What these studies actually find are residual correlations in the scaling
  relations that are in some way related to the the compactness of a
  bulge/spheroid at a given \mbh. This can be effective radius or binding
  energy or any other quantity that is ex- or implicitly a measure of
  compactness. Since bulge mass depends just linearly on the progenitor mass
  ratios, the exact size, compactness, binding energy of a spheroid produced
  in a merger will depend non-linearly on other parameters like gas fraction,
  merger orbit, etc. These parameters could easily be responsible for the
  residual correlation found in the \mm-relation as they are measures of the
  specific short- or mid-term merger history of each galaxy, while \mbulge\ is
  a long term integral. Our deliberately simple toy model itself can not make
  any statements on this issue as it does not trace e.g.\ galaxy scale radii.
  \medskip

  Interpreting the above points with respect to the relevance of AGN feedback,
  we find no strong argument for AGN feedback as a {\em necessary} mechanism
  at work. However this does not mean that AGN feedback does not exist. It
  only means that AGN feedback is still a {\em possible} mechanism involved in
  parts of galaxy evolution, which might be important e.g.\ for subclasses of
  the galaxy population\footnote{
  Undoubtedly, at least ``radio mode'' feedback \citep{crot06} is actually
  observed in some massive clusters \citep[e.g.][]{fabi06,best06} and
  possibly on the group-level \citep{giod10}.}.
 This implies that other proposed means of energy (or momentum) injection
 that have been proposed to quench SF in massive galaxies
 \citep[e.g.][]{deke06,khoc08,lofa09} are viable options. In other words,
 all other mechanisms that can be invoked to truncate star formation in
 massive galaxies appear to be equally justified.

}

\subsection{The comparison with previous work}
{
Our result is different to all previous studies, and with more than a decade
of semianalytic models including BHs having passed, one of the obvious
questions is: Why have others not found this before? The closest studies to
ours is a short proceedings contribution by \citet{gask10}, which is basically
paraphrasing \citet{peng07}, and the work by \citet{hirc10}. Their study is a
very systematic assessment of how the scaling relation scatter evolves under
the influence of galaxy merging. However, their initial setup in all cases was
an existing correlation of \mbh\ and \ms, and due to their specific goal they
deliberately ignored the influence from SF and BH accretion.

Other recent studies investigating this subject either explicitly include AGN
feedback \citep[e.g.][]{crot06,robe06,boot09,joha09,shan09} or couple the
merging scenario with a self regulation prescription for BH growth
\citep[e.g.][]{kauf00,volo09}. In retrospect it is quite obvious why
they did not find our results earlier, as most of these models were developed
for galaxy evolution in general. Once BHs entered the equations, they added
terms of (regulated) BH growth and, when evaluating the generated
\mm-relations, noted that their prescription managed to produce a decent match
to the empirical relation. The interpretation that this meant the AGN
feedback- or coupling-prescription were correct, now requires re-evaluation
from our new point of view -- it was the simple result of hierarchical
assembly at work.
}

\subsection{Impact on evolution studies}
In the last decade a substantial number of attempts were made to measure
local and higher redshift scaling relations of \mm, \msig, or
\ml\ \citep{treu04, peng06a, peng06b, woo06, treu07, schr08, jahn09b,
  merl10, benn10, deca10b} and to interprete them with respect to the
mechanisms that couple BH growth and their impact on galaxy
formation. Which implications does the non-causal origin of the scaling
relations have for these results? If the mean cosmic \ms\ and
\mbh\ actually evolved similarly, our results explain at least a part of
the {\em bulge} scaling relation evolution for galaxies with substantial
disk components: it is the simple conversion of disk to bulge mass in
galaxy mergers. What still remains interesting and needs to be
substantiated is how at high redshifts the relation between \mbh\ and
total \ms\ (or even \mbulge\ for bulge dominated galaxies) evolves
\citep[e.g.][]{walt04}. This would continue to predict a substantial early
BH growth -- with corresponding implications for BH feeding models.

One other aspect that could serve as a diagnostic is the evolution of the
scaling relation scatter. When coupled with predictions of BH and stellar
mass assembly from a proper model, the scatter can be used to study
e.g.\ the distribution of seed \mbh\ at early times.  We will follow up on
this issue in a future publication.


\acknowledgements{
  KJ is funded through the Emmy Noether Programme of the German Science
  Foundation (DFG). The authors thank F.\ Fontanot for his help in creating
  the {\sc pinocchio} merger trees, H.-W.\ Rix, E.~F. Bell, F.\ Walter, R.\
  Decarli and C.~Y.\ Peng for valuable feedback, and L.\ Mancini for access to
  their data.  Numerical simulations were performed on the PIA and PanStarrs2
  clusters of the MPIA at the MPG Rechenzentrum in Garching. We thank the
  anonymous referee for a very thorough job and very helpful comments and
  suggestions.
}



\appendix

 \renewcommand{\thefigure}{A.\arabic{figure}}
 \setcounter{figure}{0}  

\section{Testing the model: Are the parameter choices special?}
\label{sec:test}

\begin{deluxetable}{ccccclc}
\tablecolumns{7} \tablewidth{0pc} \tablecaption {Model parameters}
\tablehead{ 
\colhead{Model\tablenotemark{a}} & 
\colhead{SFR(z)\tablenotemark{b}} & 
\colhead{q\tablenotemark{c}} &
\colhead{AGN$_L$(z)\tablenotemark{d}} & 
\colhead{Stochasticity\tablenotemark{e}}& 
\colhead{B/D conversion\tablenotemark{f}} & 
\colhead{Dynamical Friction\tablenotemark{g}}
}
\startdata
A & HB06 & 0.8 & H07 & Y & M$_{G1}$/M$_{G2}$ & B08 \\
B & {\bf const.} & 0.8 & H07 & Y & M$_{G1}$/M$_{G2}$  & B08 \\
C &  {\bf const.} & 0.8 &  {\bf const.} & Y & M$_{G1}$/M$_{G2}$  & B08 \\
D & HB06 & 0.8 & H07 & {\bf N} & M$_{G1}$/M$_{G2}$  & B08 \\
E & HB06 & 0.8 & H07 & Y & M$_{G1}$/M$_{G2}$  & {\bf B08 $\times 5$} \\
F & HB06 & 0.8 & H07 & Y & {\bf (M$_{G1}$/M$_{G2}$)$^{1/2}$ } & B08 \\
G & HB06 & 0.8 & H07 & Y & {\bf (M$_{G1}$/M$_{G2}$)$^2$ }  & B08 \\
H & HB06 & {\bf 1.0} & H07 & Y & M$_{G1}$/M$_{G2}$  & B08 \\
I & HB06 & {\bf 2.0} & H07 & Y & M$_{G1}$/M$_{G2}$  & B08 \\
L & HB06 & {\bf 3.5} & H07 & Y & M$_{G1}$/M$_{G2}$  & B08 \\
\enddata

\tablecomments{Parameter of reference (A) and test models (B--L). Values in
  bold face are differences to the reference model.}
\tablenotetext{a}{Results shown in Figures~\ref{fig:modelAD}, \ref{fig:modelAG} and \ref{fig:modelAL}}
\tablenotetext{b}{Redshift dependent SFR; HB06 refers to \citet{hopk06}}
\tablenotetext{c}{Exponent of mass dependence in SFR (see Eq.~\ref{eq:sfr2})}
\tablenotetext{d}{Redshift dependent AGN luminosity; H07 refers to
  \citet{hopk07}, Figure 8}
\tablenotetext{e}{Stochastic element in BH growth}
\tablenotetext{f}{Amount of dynamical friction in galaxy mergers; B08 refers
  to \citet{boyl08}}

\label{tab:mods}
\end{deluxetable}

\begin{figure}
\begin{center}
\includegraphics[width=0.8\textwidth]{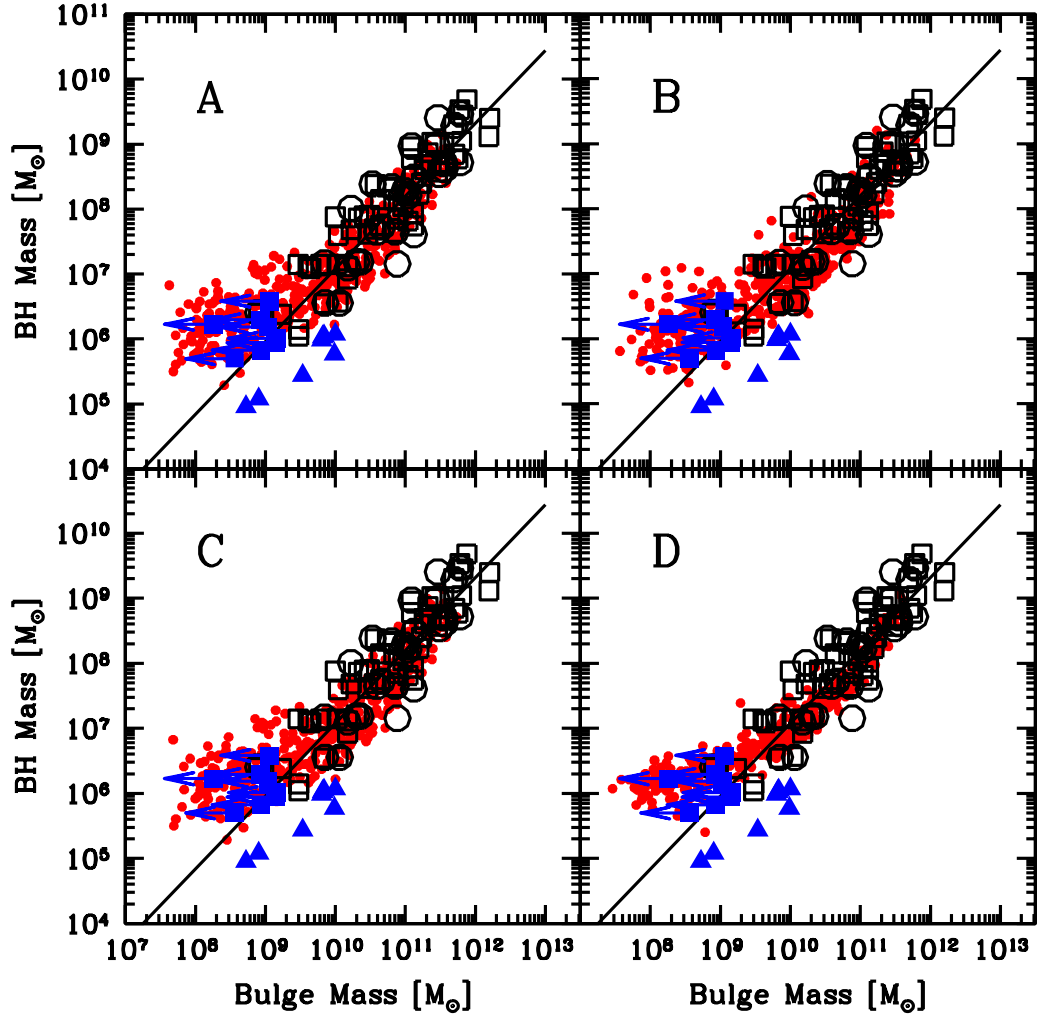}
\end{center}
\caption{Models A--D; see Table~\ref{tab:mods} for model parameters.}
\label{fig:modelAD} 
\end{figure}

\begin{figure}
\begin{center}
\includegraphics[width=0.8\textwidth]{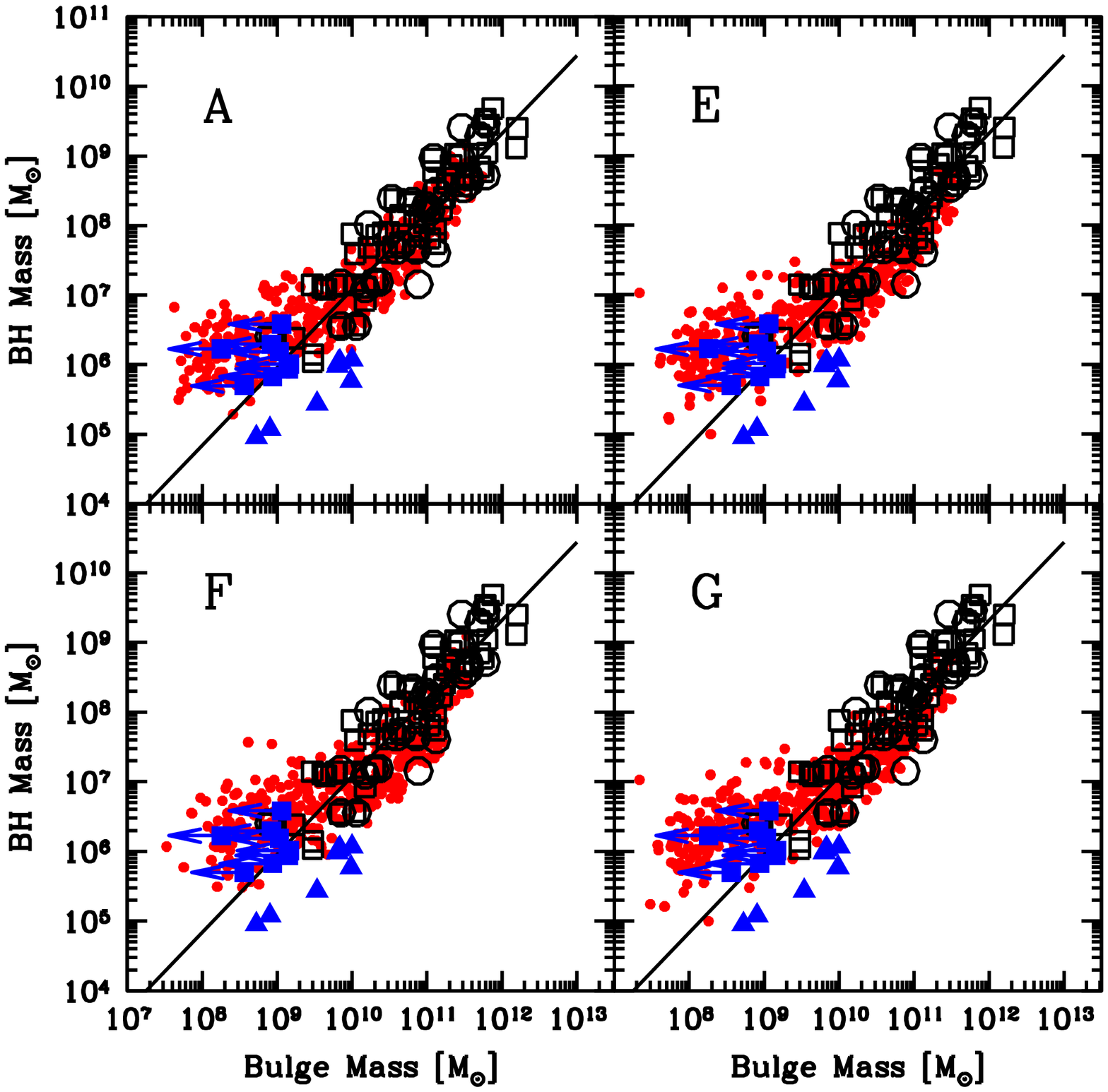}
\end{center}
\caption{Models E--G; see Table~\ref{tab:mods} for model parameters. The
  reference model A is shown again as a comparison.}
\label{fig:modelAG} 
\end{figure}

\begin{figure}
\begin{center}
\includegraphics[width=0.8\textwidth]{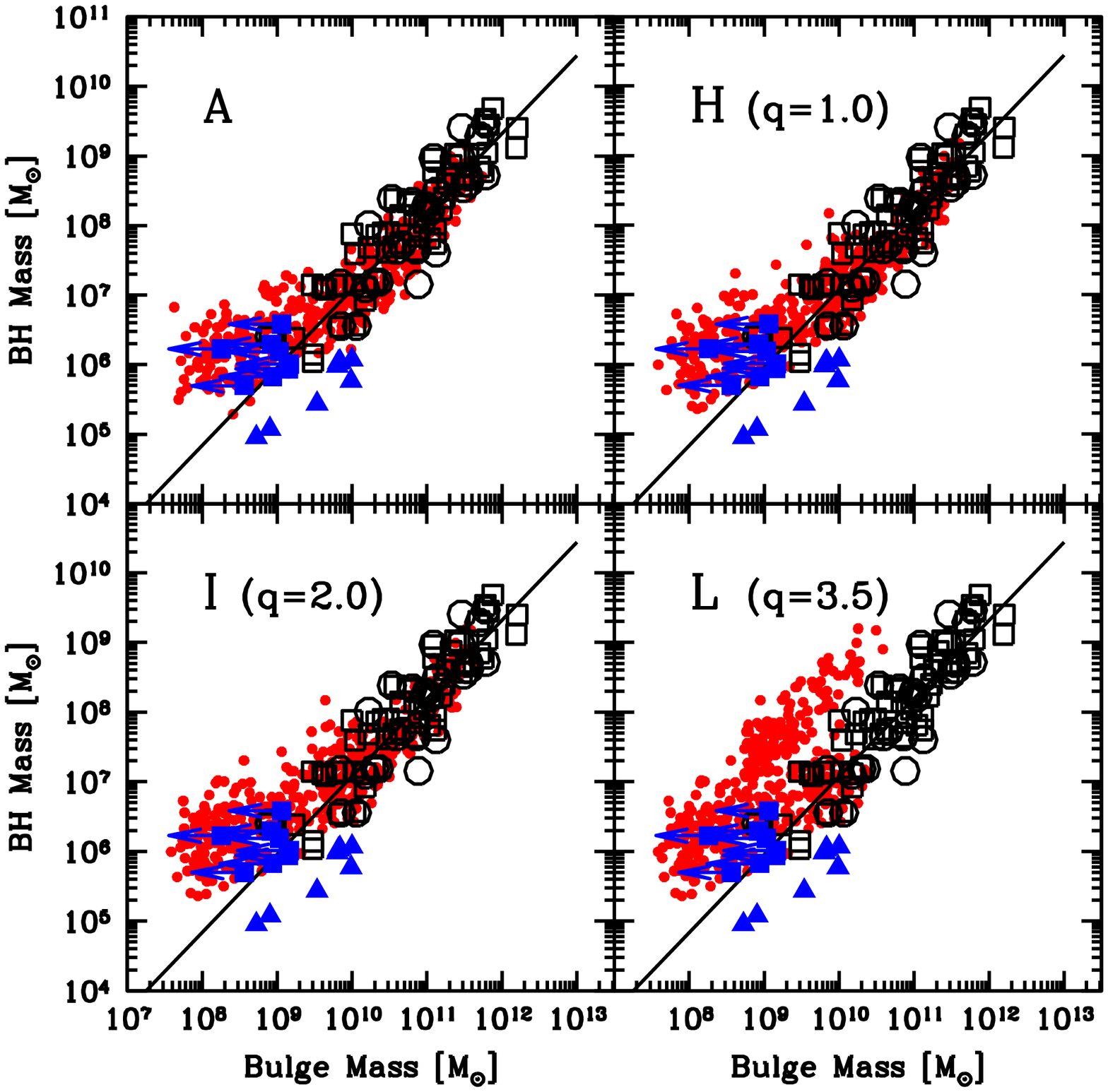}
\end{center}
\caption{Models H--L; see Table~\ref{tab:mods} for model parameters. The
  reference model A is shown again as a comparison.}
\label{fig:modelAL} 
\end{figure}

{ Few (free) parameters enter in our parametrization of star formation,
  black hole accretion and bulge to disk conversion. In this section we want
  to explore different choices with respect to our reference model and test
  their impact on the final results.  All different models are listed in
  Table~\ref{tab:mods}.

  Figure~\ref{fig:modelAD} shows the effect of varying our `weighting'
  functions.  In model B we set $SFR(z)=1.0$, e.g.\ we assume a constant star
  formation rate as a function of redshift; in model C we set also
  $AGN_L(z)=1.0$. The resulting \mm-relations are indistinguishable from the
  original (A) model. The bottom right panel of figure \ref{fig:modelAD}
  (model D) shows an even tighter correlation between \mm with respect to our
  reference model. This is because in model D we remove the stochasticity in
  the BH mass doubling, forcing all BHs to double their mass every $\tau$
  Gyrs. This increases the number of doublings in the merger tree branches
  with short life time, increasing the fraction of BH mass that is accreted
  through mergers and hence is subject to the central limit theorem.

  In Figure~\ref{fig:modelAG} we check the effect of our parametrization of
  dynamical friction (model E, where the dynamical friction time is multiplied
  by five) and of our disk-to-bulge conversion, assuming that a fraction of
  the disk mass proportional to the square-root (model F) or proportional to
  the square of the merger ratio is promoted into the disk (model G).

  Finally, models H--L (Figure~\ref{fig:modelAL}) test the importance of a
  dependence of SFR on stellar mass.  In these models we vary the $q$ exponent
  in equation~\ref{eq:sfr2}: models H and I do not show any appreciable
  variation when compared with model A. Model L which has an absolutely
  unrealistic value for $q$ is the only model where we were able to break the
  $M_{bulge}-M_{BH}$ correlation. This result can be easily understand in the
  following way: if star formation is a too strong function of stellar mass,
  then the vast majority of stars will be formed in the main branch of the
  tree that hosts (by definition) the most massive progenitor of our
  galaxy. This implies that the bulk of stellar mass will not be subject to
  any averaging process and hence the central limit theorem does not apply.
  Moreover given the artificially high fraction of stars produced within the
  central galaxy, there will be no major merger, this explains why in model L
  we do not get any bulge more massive than $4\times 10^{10}$ \msun.

  All other model are not distinguishable from the reference model A, this
  underlines one more time the convergence power of galaxy mergers and shows
  that the actual implementation of star formation, dynamical friction, BH
  accretion and bulge formation are only secondary effects.  }

\section{Scaling relations with stellar mass and halo mass}\label{sec:stellardm}
\setcounter{figure}{3}

\begin{figure*}
\includegraphics[width=0.48\textwidth]{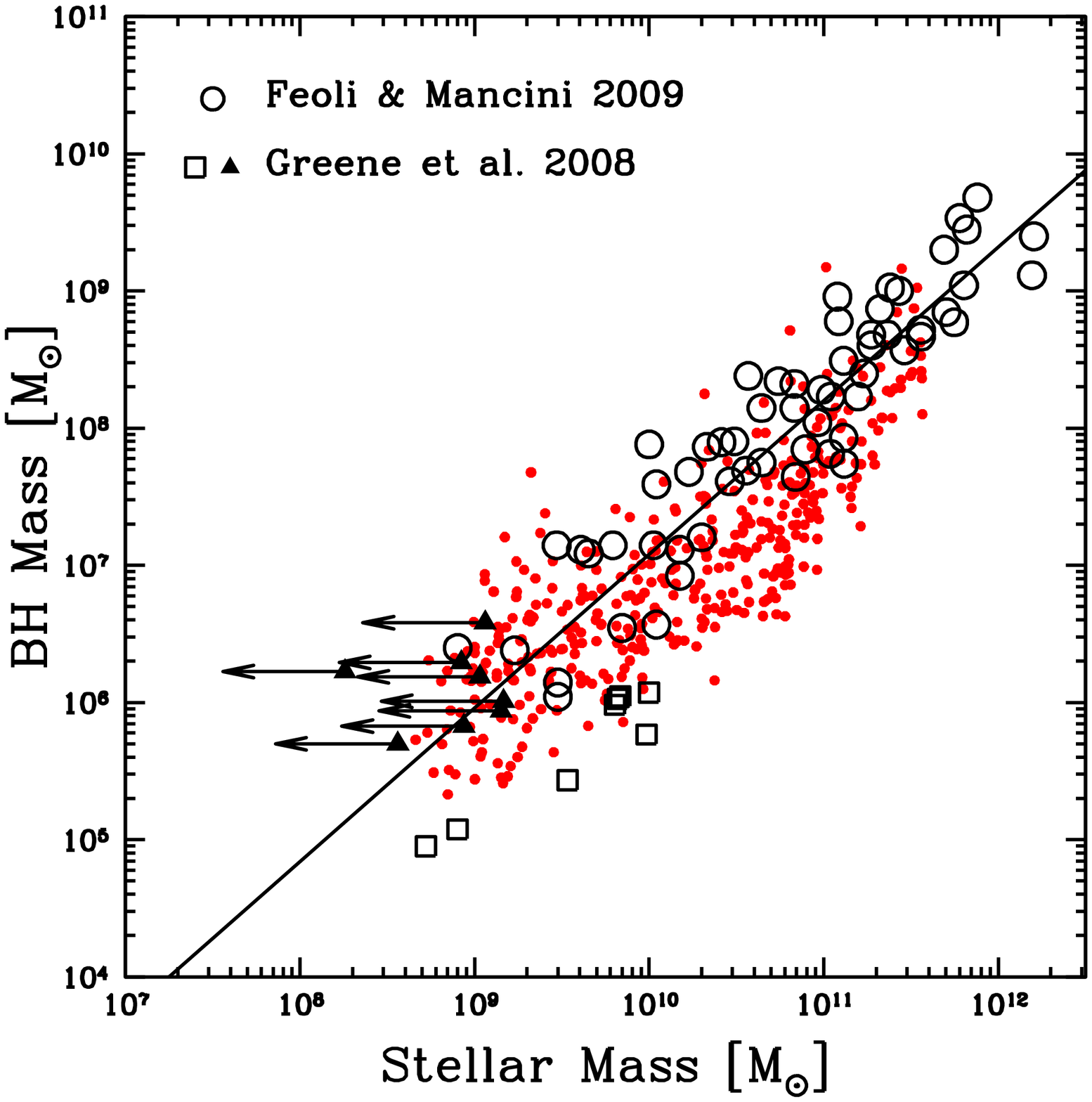}
\includegraphics[width=0.48\textwidth]{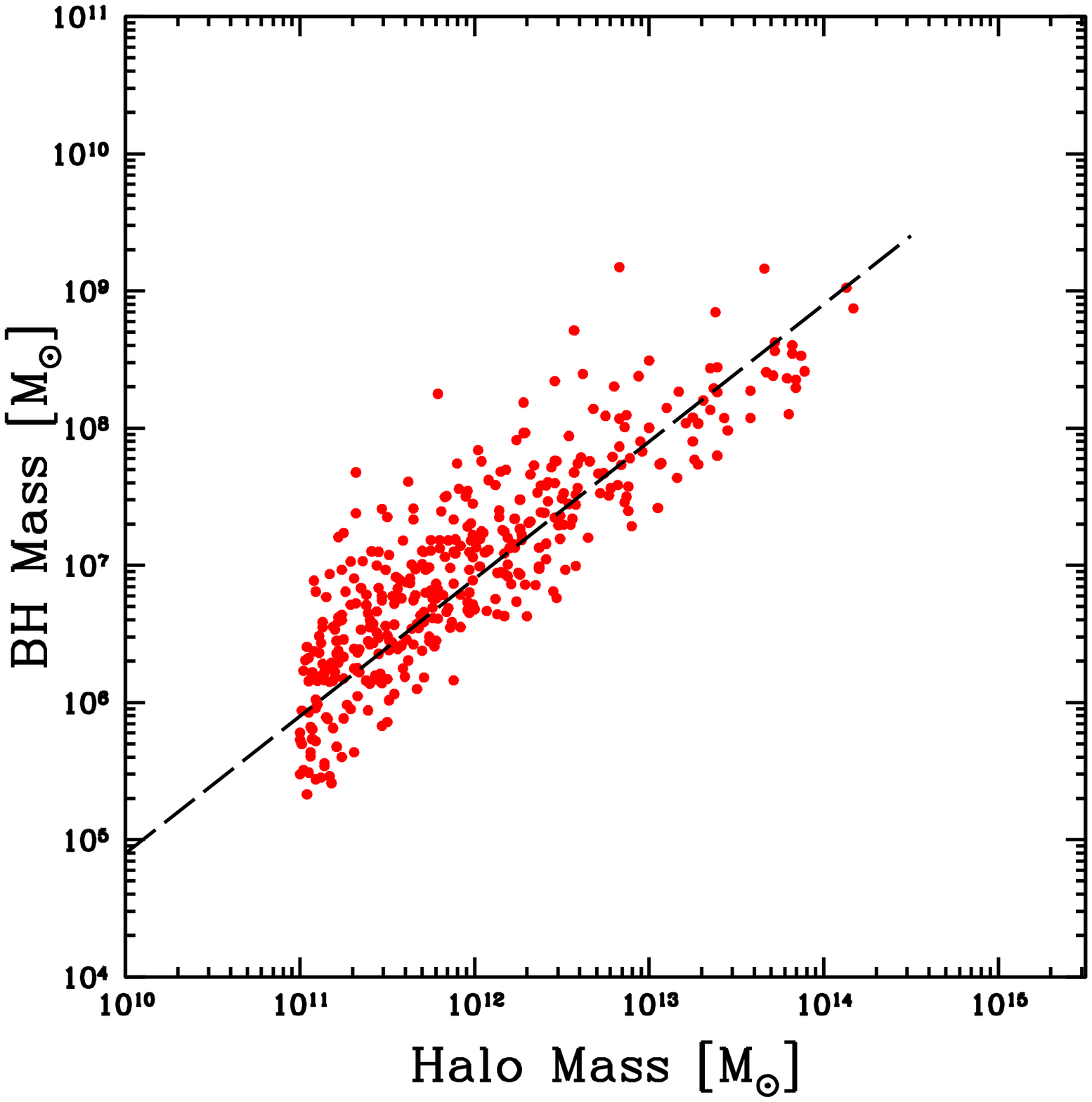}
\caption{ \label{fig:mstellar} 
 Left: Black hole vs.\ total stellar mass for 400 of our model halos (red
points), compared as in Figure~\ref{fig:MbMbh1} to the observed local
relation in black, including the compilation from \citep[circles]{feol09},
low-mass spheroids (open squares) and upper limits for spiral bulges
(triangles) from \citet{gree08}. As before, the solid line is the linear
fit by \citet{haer04} with a slope of 1.12.
Right: Black hole vs.\ dark matter halo mass. The dashed line with slope=1
is plotted to guide the eye. Please note: The abcissa is shifted by 3dex
with respect to the stellar mass plot on the left.}
\end{figure*}

{ Hierarchical assembly produces correlations between any two
  parameters that take part in this cascade. The main focus of this paper
  lies on the \mm-relation, but \mbh\ also correlates with total stellar
  mass and dark matter halo mass. This is shown in the two panels of
  Figure~\ref{fig:mstellar}, which include the receipes for BHA. The
  clearest and simplest correlation is the one with \mdm\ (right side),
  which in the simple framework of this toy model is very close to
  slope=1, with some scatter and no bending.

\mbh\ vs.\ total stellar mass (incl.\ SF receipe), as shown on the left
side of Figure~\ref{fig:mstellar} is at the massive end largely identical
to the \mm-view, because most galaxies there will have had sufficient
numbers of minor and major mergers in order to make them bulge
dominated. At the low-mass end, the scatter is still larger than at high
masses, due to the smaller number of (averaging) past mergers for each
halo, but it is somewhat smaller than for bulge mass, since the extra
random element from disk-to-bulge conversion is absent. This also explains
the missing plume of seemingly high \mbh-systems at low masses visible in
Figure~\ref{fig:MbMbh1}, which in this way can be explained to actually be
disk galaxies with small bulges and normal sized black holes for the
amount of total stellar mass.}

\end{document}